\documentclass[%
 aip,
 amsmath,amssymb,
reprint,%
]{revtex4-1}
\usepackage[dvipdfmx]{graphicx}
\usepackage{epsfig}

\newcommand{\ra}{\rangle}
\newcommand{\la}{\langle}

\begin{document}
\title{Robustness of cluster states and surface code states against random local fields}
\author{T. Tanamoto}
\affiliation{Department of Information and Electronic Engineering, Teikyo University,
Toyosatodai, Utsunomiya  320-8511, Japan} 

\author{M. Ueda}
\affiliation{Department of Physics, University of Tokyo, Hongo, Bunkyo-ku, Tokyo 113-8654, Japan}

\begin{abstract}
In ideal quantum circuits,  qubits are tacitly assumed to be uniformly fabricated and operated 
by prescribed signals.
In reality, however, we must cope with different control signals to adjust individual qubits, 
which requires a large overhead of control circuits.
Here, we theoretically investigate how random local fields affect cluster states and surface code states which constitute the key highly entangled states in quantum computation. 
We find similar behavior of temporal degradation of the fidelity 
for both cluster states and surface code states for the number of qubits up to ten.
We find that the effect of local field fluctuations is greatly mitigated if the magnitude of fluctuations can be suppressed below 10\% of the energy gap $\Delta$
for both cluster states and surface code states. 
If the magnitude of fluctuations exceeds $\Delta/2$, 
the fidelity for both states  deteriorates dramatically.
A simple estimation based on the average fidelity up to $t\sim 2\hbar/\Delta$ shows that  
the maximum number of qubits  
that can be corrected with the 1\% error threshold is less than 31 
for surface code states and 27 for cluster states.
This means that the 
error correction should be carried out during a time shorter than $2\hbar/\Delta$.
\end{abstract}
\maketitle

\section{Introduction}
Quantum computation has seen a surge of interest since quantum annealing machines 
based on superconducting qubits were commercialized by D-wave systems Inc.~\cite{D-wave}. 
Moreover, IBM has made small-size quantum 
computers available to wide users through internet connection~\cite{IBM}.
The number of qubits in quantum computing devices is currently
below one hundred, but it is expected to increase significantly in the years to follow.
Thus, it will become increasingly more important to investigate quantum computing systems 
from a viewpoint of engineering reliability as in semiconductor devices.
It seems not long before quantum computing systems can be prepared with near-unity fidelity; then we will face the problem of how robustly they can operate against local field fluctuations caused by defects and ambient electromagnetic fields
due to crosstalk, wiring, etc.
It is therefore important to examine the effect of 
local field fluctuations on the state of a system 
whose number of qubits exceeds 50
where the quantum advantage becomes experimentally feasible~\cite{Boixo,Villalonga}.

Here, we numerically investigate the effect of unavoidable local 
field fluctuations on otherwise perfect qubit states.
In integrated circuits, qubits are supposed to operate uniformly
where qubits change their quantum state with some prescribed signals.
If there are additional local fields, 
qubit states change from their ideal points, 
resulting in errors.
Thus, the additional local fields require an additional overhead 
to the circuits.
Even after the technologies of qubits advances, 
it will remain a major challenge to perfectly control the fabrication process of the qubit system and realize the perfect qubit system~\cite{Queisser}.
Besides, there are ambient electromagnetic fields ranging from low to high frequencies.
Unless the electric circuits are not shielded from external electromagnetic environments, 
they suffer electromagnetic noises caused by unavoidable defects or trap sites.

At present, solid-state qubits are fabricated based on materials
such as Si, SiGe/Si, GaAs/AlGaAs and superconducting materials.
Table 1 lists the defect density and the defect-free area for solid-state qubits.
Let $n_{\rm imp}$ cm${}^{-2}$ be the number density of defects. 
If the area $S$ of a qubit is larger than the $S_{c}\equiv 1/n_{\rm imp}$,
it has on average more than one defect.
The bottom row of Table 1 shows the maximum defect-free area in units of 100 nm${}^2$.
For example, a Si-qubit with an area larger than 100 nm $\times$ 100 nm, 
has a significant probability of including one defect in its conducting area. 
The integrated circuits which include thousands of qubits 
will inevitably involve numerous defects and suffer fluctuations caused by them.
This presents a major challenge for solid-state quantum computers which require highly entangled states for information processes.
Thus, it is vital to investigate the effect of the local fluctuating fields on 
qubit systems for near-future quantum computation. 

\begin{table}
\begin{center}
Table 1: Defect density and the defect-free device area for solid-state qubits.
Data are taken from Ref.~\cite{Bedell} (SiGe/Si), Ref.~\cite{Green} (Si/SiO${}_2$), Ref.~\cite{Magno} (GaAs/AlGaAs), and Ref.~\cite{Stoutimore} (Al/AlO${}_x$).
\begin{tabular}{|l|c|c|c|c|}\hline\hline
Material &
SiGe/Si & Si/SiO${}_2$ &  GaAs/AlGaAs & 
 Al/AlO${}_{x}$ \\ 
\hline 
\begin{tabular}{l}
defect density\\$[$cm${}^{-2}]$
\end{tabular} &
$\sim 10^{8}$  & $\sim 10^{10}$ & $\sim 10^{11}$ & $\sim 10^{4}$ \\
\hline
\begin{tabular}{l}
defect-free \\ area /(100 nm${}^{2}$) 
\end{tabular} &
$\sim 10^{4}$  & $\sim 10^{2}$ & $\sim 10 $ & $\sim 10^{8}$  \\
\hline\hline
\end{tabular}
\end{center}
\end{table}

The cluster state is a highly entangled state 
in the measurement-based quantum computation~\cite{Briegel,Hein}
in which a cluster state is initially prepared and then 
quantum computation is carried out through measurement of qubits one by one.
Experiments have been carried out mainly in optical systems~\cite{Zeilinger,Plenio,Gao}
such as photonic cluster state systems using a series of 
emitted photons~\cite{Rudolph1,Rudolph2,Rudolph3} and
continuous variable cluster states~\cite{Menicucci,Moore}.
One of the authors (T.T.) has proposed how to produce cluster states 
from solid-state qubits~\cite{tanaPRL,tanaJJAP}.
In any qubit system, it will not be practical to ignore
unexpected local fluctuations 
because we cannot prepare a perfect qubit lattice.
The surface code has been studied by a number of researches 
both theoretically~\cite{Kitaev1,Kitaev2,Fowler,Fowler2} and experimentally~\cite{Xmon2,Xmon3}.
The qubit system consists of $data$ qubits and $measurement$ qubits 
where the data qubits form a logical qubit state and errors are detected and corrected by the measurement qubits.
In the standard setup, the measurement qubits are placed close to the data qubits.
In general, it is assumed that qubit-qubit interactions 
are switched on when necessary.
Because a number of operations have to be applied to the measurement qubits next to 
the data qubits, 
unexpected dynamical noises affect the data qubits. 
Thus, it is important to consider the effect of local random fields 
on the qubit state of the surface code.

The standard approach to the analysis of the decoherence and degradation of 
qubits is to use master equations~\cite{Hutter,Freeman,Viyuela,Jennings} 
which are very effective to describe ensemble-averaged characteristics of the qubit system. 
However, it is not easy to include local field fluctuations. 
There are other sophisticated approaches to understand the noise properties~\cite{Cirac,Hamma,Dusuel,Amaro}.
Here, we take a direct and simple approach to local fluctuations.
We numerically investigate the effect of local field fluctuations 
caused by internal and external electromagnetic fields by 
adding the corresponding terms to an ideal Hamiltonian.  
We describe how the fidelity deteriorates as the number of fluctuating qubits and the magnitude of fluctuations increase. 
Specifically, we assume an ideal cluster state or a surface code state initially,
and numerically solve the Schr\"{o}dinger equation of the Hamiltonian $H_0+H_1$, 
where $H_0$ describes an ideal cluster or surface code state, and 
$H_1$ describes the effect of local field fluctuations. 
Fowler {\it et al}.\cite{Fowler2} show that a 1 \% error can be corrected by the quantum error correction in the surface code.
Thus, we regard the 1 \% degradation of the fidelity as the threshold 
above which fidelity can be improved by the standard quantum error-corrections.

Because of the constraint of our computer resources, 
the number of qubits is limited up to $N=8$ for the cluster state 
and $N=9$ for the surface-code Hamiltonian.
In this paper, we consider the effect of local fluctuations 
by examining the time-dependent solutions of the Schr\"{o}dinger equations. 
In the first part, we add the fluctuating local fields to all qubits.
We show that when the fluctuations are small,
the fidelity of wave functions remains close to unity.
We find that for both cluster states and surface code states
the effects of high-frequency local 
fields are similar to those of static local fields.
We also investigate a situation in which fluctuating fields 
are added to some part of qubits 
to understand the relationship between the 
number of qubits with the local fluctuations and the fidelity.
Because both cluster states and surface code states are 
highly entangled, 
one might expect that a single local fluctuating field drastically 
changes the fidelity and the magnitude of the local fluctuating
fields are of secondary importance to the fidelity. 
However, our numerical result shows that the degradation of the fidelity
depends critically on it and, furthermore, it is proportional to the number of the qubits 
in the presence of fluctuating fields.
This is similar to the classical devices:
the reduction in the number of noise sources directly 
improves the device performance.

Because there is no dissipation in the present framework,
the wave functions is a periodic function of time.
When the coherence time is short compared 
with the period of local field fluctuations, 
we will have to consider the effect of 
dissipation in addition to the random qubit system.
In this paper, we take into account the effect of dissipation 
by introducing an optical potential~\cite{Zohta}   
which is a non-Hermitian term 
added to the original Hamiltonian to describe various decay 
phenomena~\cite{Ueda}.

The rest of this paper is organized as follows.
In Sec.~\ref{sec:formalism}, we briefly review the cluster state and the surface code state. 
We also discuss the cases of two qubits that can be solved analytically. 
In Sec.~\ref{sec:results}, we show our numerical results 
about the effects of locally random fluctuations.
In Sec.~\ref{sec:optical}, we show numerical results for the effect of the optical potential.
In Sec.~\ref{sec:discussion}, we briefly discuss the origin of the local fluctuation term, 
 and estimate the order of an operation time discussed in this paper.
In Sec.~\ref{sec:conclusion}, we provide a summary and conclusions of this paper.

\begin{figure}[h]
\begin{center}
\includegraphics[width=4cm]{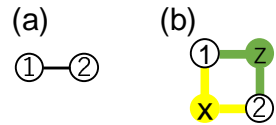}
\end{center}
\caption{Minimal cluster (a) and surface code (b) constituted from qubits. 
The while circles show the data qubits 
and the colored ones show the measurement qubits,
where $X$ and $Z$ denote the $x$ and $z$ components of the Pauli matrices. }
\end{figure}

\section{Formulation}\label{sec:formalism}
The Hamiltonian of the system is given by $H=H_0^{\alpha}+H_1$ 
($\alpha$=c,s) where the superscript $\alpha=c$ and $s$ refer to
the cluster state and the surface code state, respectively, and
$H_1$ expresses the effects of local field fluctuations given by
\begin{equation}
H_1=\sum_{i=1}^N [g_i(t) X_i+h_i(t) Z_i],
\end{equation}
where $N$ is the number of qubits, $g_i(t)= g_i \cos( \omega_i^x t)$
and $h_i(t)= h_i \cos( \omega_i^z t)$ are randomly oscillating fields, and 
$X_i=\left( \begin{array}{cc} 
0 &1 \\
1& 0
\end{array}\right)$
 and 
 $Z_i=\left( \begin{array}{cc} 
1 & 0 \\
0 & -1
\end{array}\right)$ are the $x$ and $z$ components of the Pauli matrices, respectively.
The amplitudes ($g_i$ and $h_i$) and the frequencies 
($\omega^x_i$ and $\omega^z_i$) are chosen from random numbers such that $  |h_i(t)| \le h_{\rm ave}$,  
$|g_i(t)| \le g_{\rm ave}$, $|\omega^x_i| \le \omega^x_{\rm ave}$,
 and $|\omega^z_i| \le \omega^z_{\rm ave}$ 
 for given $h_{\rm ave}$, $g_{\rm ave}$, $\omega^x_{\rm ave}$,
 and $\omega^z_{\rm ave}$. 
To extract the effect of the local fluctuation, 
dissipation is not included at this level.
Thus, the effect of the additional term alters the phase of the 
ideal wave function, causing the fidelity of the system 
to decrease from unity.

\subsection{Cluster state}
As described in Ref.\cite{Hein}, 
it is straightforward to analyze the eigenstates of cluster states. The cluster state $|\Phi_C\ra $ is an eigenstate of the stabilizer operator
\begin{equation}
K_i= X_i \underset{j \in {\rm nbhd}(i)}{\Pi} Z_j,
\end{equation}
such that $K_i|\Phi^C\ra=(-)^{\kappa_i}|\Phi^C\ra$ ($\kappa_i=\{0,1\}$), 
where nbhd($i$) denotes the nearest neighborhood of qubit $i$. 
The corresponding cluster-state Hamiltonian is given by
\begin{equation}
H_0^{c}=-\Delta \sum_i K_i,
\end{equation}
where $\Delta$ characterizes the energy gap of cluster states.
We consider the lowest-lying cluster state, $|\Phi^C_0\ra$,
such as $K_i|\Phi^C_0\ra=|\Phi^C_0\ra$.

In the cluster states, the excited states are determined from 
the action of $Z_i$ on the ground state~\cite{Hein}. 
Because $
H_0^c|\Phi^C_0\ra /\Delta =-\sum_{j} K_j |\Phi^C_0 \ra=-N|\Phi^C_0\ra$,
we have
\begin{eqnarray}
\lefteqn{
HZ_i |\Phi^C\ra /\Delta =-( K_i +\sum_{j \ne i} K_j )|\Phi^C\ra 
}\nonumber \\
&=& -Z_i(- K_i +N-1)|\Phi^C\ra
=-Z_i(N-2)|\Phi^C\ra.
\end{eqnarray}
The second excited states are given by $Z_iZ_j|\Phi^C\ra$ and
similarly the $n$-th excited states are given 
by $\Pi_{i=1}^n Z_i |\Phi^C\ra$ $(n\le  N)$.
For $N=4$, there is a unique ground state 
of energy $-4\Delta$, and two first excited states 
of energy $-2\Delta$, the next excited state of energy $2\Delta$, 
and the highest excited state of energy $4\Delta$.
Thus, we can explicitly write down the 
matrix elements for the fluctuating term $\sum_i h_i Z_i$
by expressing the Hilbert space in terms of $\Pi_{i=1}^n Z_i |\Phi^C\ra$.
We can also explicitly write down the matrix elements 
for the fluctuating term $\Pi_{i=1}^n X_i |\Phi^C\ra$ by using the 
characteristics of the stabilizer code.
For the case of $N=4$, we have 
\begin{equation}
X_1 |\Psi_0^C\ra = X_1 (X_1Z_2Z_4)|\Psi_0^C\ra=Z_2Z_4|\Psi_0^C\ra.
\end{equation}
Thus, the effect of the $x$-perturbation brings each state into two higher eigenstates.

\subsection{Surface code}
\begin{figure}[h]
\begin{center}
\includegraphics[width=8cm]{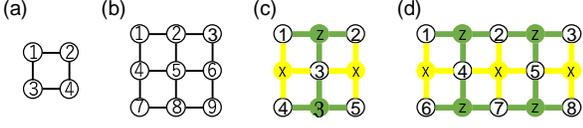}
\end{center}
\caption{Clusters with (a) $N=4$ and (b) $N=9$, and 
surface codes with (c) $N=5$ and (d) $N=8$.}
\label{Fig2}
\end{figure}

The surface code, which is defined as a qubit system on a planar square lattice, 
is one of the stabilizer codes
which are simultaneous eigenspaces of check operators.
The check operators of the surface code consist 
of products of Pauli matrices $X_i$ and $Z_i$, and 
the corresponding Hamiltonian is given by
\begin{equation}
H_0^{s}=-\Delta \sum_i [\sum_{l\in {\rm star}(i)}  \Pi_l X_{l} 
+\sum_{l'\in {\rm boundary}(i)}  \Pi_{l'}Z_{l'}].
\label{surface}
\end{equation}
A realistic qubit system for the surface code consists of 
{\it data} qubits and {\it measurement} qubits~\cite{Fowler}.
The data qubits constitute the logical system of Eq.~(\ref{surface}),  
and the measurements qubits play the role of error-correcting 
operations on the data qubits. 
Here we identify the number of the data qubits as the number of qubits $N$
in our calculations.
The minimum surface code includes four qubits, two of which are 
data qubits ($N$=2) and the other two are measurement qubits that control the logical qubits (Fig.~1(b)).
The next larger system includes five data qubits ($N=5$), 
and four measurement qubits as shown in Fig.~\ref{Fig2}(c).
The $N=5$ surface-code Hamiltonian is given by
\begin{equation}
H_0^{s}=-\Delta[X_1X_3X_4+X_2X_3X_5+Z_1Z_2Z_3+Z_3Z_4Z_5].
\end{equation}
The ground state of the surface code is nondegenerate, 
and the eigenenergy of the $N=5$ system is $-4\Delta$. 
The eigenenergies of the exited states are 
given by $-2\Delta$, 0, $2\Delta$, and $4\Delta$. 
For the surface-code Hamiltonian, there is no useful formula unlike for the cluster states, 
and we directly construct the system Hamiltonian and solve eigenvalue problems numerically.
Note that, to realistically correct quantum errors,
the minimum size of the surface code is $N=25$~\cite{Fowler}.
However, because of the limitation of computational resources, 
we consider a qubit system for $N \le 8$ surface codes.

\subsection{Time-dependent fidelity}
We numerically solve the Schr\"{o}dinger equation
\begin{equation}
i\hbar \frac{\partial |\Phi\ra}{\partial t}
=H|\Phi\ra
\end{equation}
by using the Runge-Kutta method in the $Lapack$ subroutines.
We consider three types of qubit configurations for 
the cluster state and the surface code state as shown in Fig.~1. 
For cluster states, we consider $N=2,4,9$ qubits,  
and, for surface code states, we consider $N=2,5, 8$ qubits.

The fidelity is defined as
\begin{equation}
F(t)= |\la \Phi_0 (t) | \Phi (t) \ra |^2, 
\end{equation}
where $|\Psi_0 (t) \ra$ is a time-dependent wave function 
of $H_0$.

\begin{figure}
\begin{center}
\includegraphics[width=8.5cm]{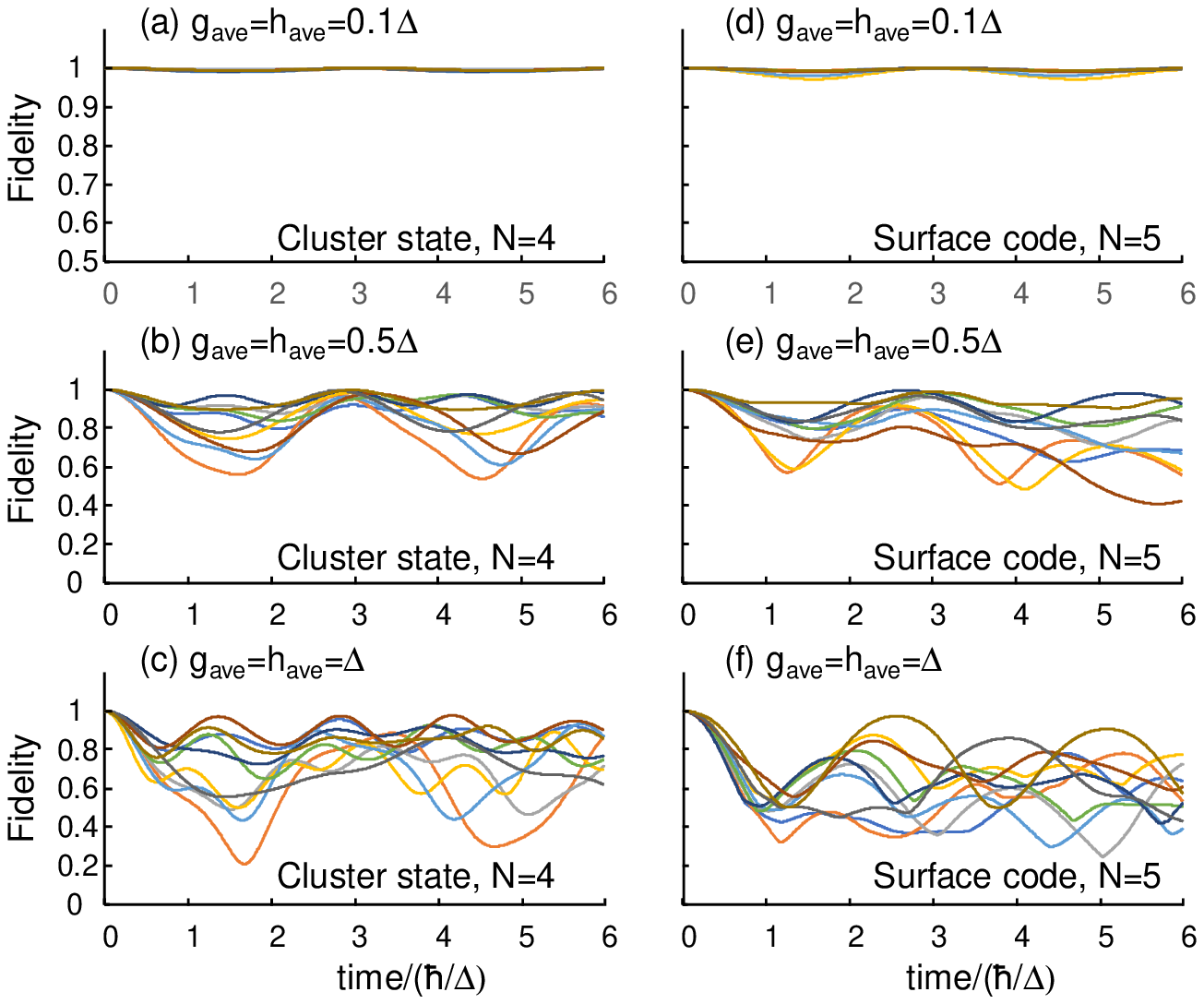}
\end{center}
\caption{Time evolution of the fidelity for the $N=4$ cluster state (a-c) and 
that of the $N=5$ surface code state (d-f) under static local fields $\omega^x_i=\omega^z_i=0$, 
where 
$g_{\rm ave}=h_{\rm ave}=0.1\Delta$ for (a) and (d), 
$g_{\rm ave}=h_{\rm ave}=0.5\Delta$ for (b) and (e), and
$g_{\rm ave}=h_{\rm ave}=\Delta$ for (c) and (f). }
\label{FigCal1}
\end{figure}
\begin{figure}
\begin{center}
\includegraphics[width=8.5cm]{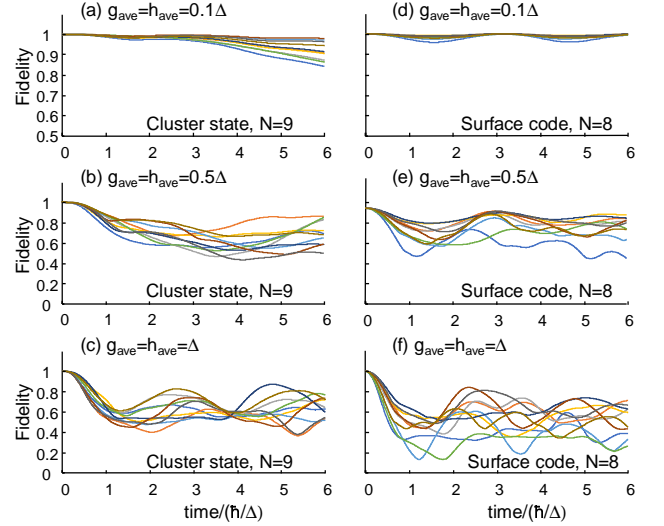}
\end{center}
\caption{Time evolution of the fidelity for the $N=9$ cluster state (a-c) and 
that of the $N=8$ surface code state (d-f) under static local fields $\omega^x_i=\omega^z_i=0$,
where
$g_{\rm ave}=h_{\rm ave}=0.1\Delta$ for (a) and (d), 
$g_{\rm ave}=h_{\rm ave}=0.5\Delta$ for (b) and (e), and
$g_{\rm ave}=h_{\rm ave}=\Delta$ for (c) and (f). }
\label{FigCal2}
\end{figure}

\subsection{Minimum system}
Let us first
examine the analytic solutions of the minimum systems for both 
the cluster state and the surface code state.
The minimum cluster state of $N=2$ is 
given by $|\Phi_0^C\ra =(|0\ra |+\ra +|1\ra |-\ra)/\sqrt{2}$ (Fig.~1 (a)).
The ground state of the minimum surface code of $N=2$ is given by a singlet state 
of $(|0\ra |1\ra -|1\ra |0\ra)/\sqrt{2}$~\cite{Fowler}(Fig.~1 (b)).
The Hamiltonian of the minimum cluster state is given by
\begin{equation}
H_0^c=-\Delta(X_1 Z_2+Z_1X_2).
\end{equation}
The three excited states are given by 
$|\Phi^{(1)C}_L\ra=Z_1 |\Phi_0^C\ra$,
$|\Phi^{(1)C}_R\ra=Z_2 |\Phi_0^C\ra$, and
$|\Phi^{(2)C}\ra=Z_1Z_2 |\Phi_0^C\ra$, 
whose eigenenergies are, 0,0, and $2\Delta$, respectively.
The total Hamiltonian is given by
\begin{equation}
H=\left(
\begin{array}{cccc}
   -2\Delta  & h_1+g_2 & h_2+g_1 & 0       \\
h_1+g_2 &  0      & 0       & h_2-g_1 \\
h_2+g_1 &  0      & 0       & h_1-g_2 \\
0       & h_2-g_1 & h_1-g_2 & 2\Delta       
\end{array}
\right).
\end{equation}
The eigenvalues of the simplest case of $g_1=g_2=0$
is given by
\begin{equation}
E=2\Delta^2+h_1^2+h_2^2 \pm 2 \sqrt{ (\Delta^2+h_1^2)(\Delta^2+h_2^2) }.
\end{equation}
Thus, the effect of the fluctuations is  
to shift the eigenenergy when the magnitude of 
the fluctuation is small compared with $\Delta$.
The ground state is given by
\begin{equation}
a^c |\Phi_0^C\ra
+b^c |\Phi_L^C \ra
+c^c |\Phi_R^C \ra
+d^c|\Phi^{2C} \ra,
\end{equation}
where 
$a^c=(q_1+\Delta)(q_2+\Delta)/\sqrt{D^{c}}$, 
$d^c=h_1h_2/\sqrt{D^{c}}$, 
$b^c=-h_1(q_2+\Delta)/\sqrt{D^{c}}$, and
$c^c=-h_2(q_1+\Delta)/\sqrt{D^{c}}$,
using $q_i=\sqrt{\Delta^2+h_i^2}$ (i=1,2) and 
$D^{c}=(q_1+\Delta)^2(q_2+\Delta)^2+h_1^2h_2^2$. 
Thus, as the magnitude of the fluctuations $h_i$ 
increases, the wave function evolves from 
the ground state $|\Phi_0^C\ra$ to higher excited states.
 
The minimum surface-code Hamiltonian is given by
\begin{equation}
H_0^s=-\Delta (X_aX_b +Z_a Z_b).
\end{equation}
This corresponds to the XY model and the eigenstates 
are the Bell states.
The ground state is spin-singlet as mentioned above.
The total Hamiltonian including the local fluctuations 
is given by
\begin{equation}
H=\left(
\begin{array}{cccc}
 -\Delta+h_1&  g_2   & g_1 & -\Delta       \\
  g_2 &  \Delta+h_2 & -\Delta       & g_1 \\
  g_1 &  -\Delta     &  \Delta-h_2   & g_2 \\
 -\Delta   & g_1    &  g_2 & -\Delta-h_1       
\end{array}
\right).
\end{equation}
For $g_1=g_2=0$, the eigenenergies read
\begin{eqnarray}
E_-=-\Delta\pm \sqrt{\Delta^2+h_m^2}, \ 
E_+= \Delta\pm \sqrt{\Delta^2+h_p^2}, \ 
\end{eqnarray}
where $h_p\equiv h_1+h_2$ and $h_m\equiv h_1-h_2$.
Thus, the effect of the local fluctuation 
is to shift the energy.
The wave function of the ground state is given by 
$b^s |01\ra -c^s|10\ra$, where 
$b^s=\Delta/\sqrt{2\sqrt{D^s}(\sqrt{D^s}+h_m)}$
and
$c^s=(\Delta+\sqrt{D^s})/\sqrt{2\sqrt{D^s}(\sqrt{D^s}+h_m)}$
with $D^s=\Delta^2+h_m^2$.
Thus, the wave function of the surface code slightly changes from 
the original singlet state.

\section{Numerical results}\label{sec:results}
The random numbers are taken from $[-1,1]$ 
and multiplied by
$g_{\rm ave}$, $h_{\rm ave}$, $\omega^x_{\rm ave}$, and $\omega^z_{\rm ave}$.
All calculations are repeated ten times to examine the randomized effects.
Thus, each figure includes ten curves.
Figure \ref{FigCal1} shows the numerical results of the time-dependent fidelities
for $N=4$ cluster states (a-c) and $N=5$ surface code states (d-f). 
For small static fluctuations of $g_{\rm ave}=h_{\rm ave}=0.1\Delta$, 
the fidelity is stable with no large variations. 
When the magnitude of the fluctuations is half of $H_0$ ({\it i.e.}, 
$g_{\rm ave}=h_{\rm ave}=\Delta/2$) the fluctuations 
significantly disturb the original wave functions.
For $g_{\rm ave}=h_{\rm ave}=\Delta$, 
the fidelity changes substantially.
In addition, we see no pronounced distinction 
between the cluster states and the surface code states.

Figure \ref{FigCal2} shows the numerical results of the time-dependent fidelity
for $N=9$ cluster states (a-c) and $N=8$ surface code states (d-f).
For the surface code state, we can see a tendency similar to Fig.~\ref{FigCal1}.
That is, for small static fluctuations of $g_{\rm ave}=h_{\rm ave}=0.1\Delta$, 
the fidelity remains stable, 
but as the magnitude of the fluctuations is increased to$\Delta/2$, 
the modulations of wave functions become larger and larger. 
For the cluster state, the fidelity of $N=9$ deteriorates 
after $t\sim 3\Delta$. 
During $t < 3\Delta$, we can see similar behavior for the cluster state.
It is striking that the behavior of the surface code shown in Fig.~\ref{FigCal2} 
is very similar to that in Fig.~\ref{FigCal1}, 
despite the fact that the number of qubits is almost doubled. 
However, as we show later in Fig.~\ref{FigCal6}, the fidelity becomes 
smaller with increasing $N$.
The similarity between Fig.~\ref{FigCal1} and Fig.~\ref{FigCal2}
indicates that the degradation rate is not large even as the size of the system 
increases.

\begin{figure}
\begin{center}
\includegraphics[width=8.5cm]{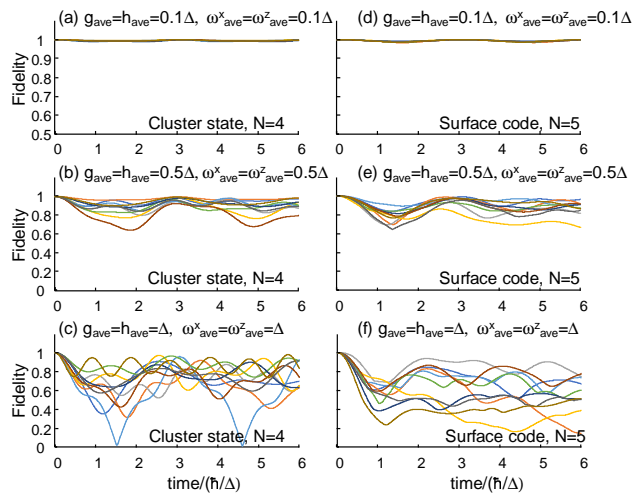}
\end{center}
\caption{Time evolution of the fidelity of the $N=4$ cluster state (a-c), and 
that of the $N=5$ surface code state (d-f) under oscillating local fields.
(a) and (d) for $g_{\rm ave}=h_{\rm ave}=0.1\Delta$,
and $\omega^x_{\rm ave}=\omega^z_{\rm ave}=0.1\Delta$. 
(b) and (e) for $g_{\rm ave}=h_{\rm ave}=0.5\Delta$,
and $\omega^x_{\rm ave}=\omega^z_{\rm ave}=0.5\Delta$. 
(c) and (f) for $g_{\rm ave}=h_{\rm ave}=\Delta$,
and $\omega^x_{\rm ave}=\omega^z_{\rm ave}=\Delta$.
For each set of parameters, ten samples are taken to examine 
the randomized effects. Each figure includes ten different results. } 
\label{FigCal3}
\end{figure}

\begin{figure}
\begin{center}
\includegraphics[width=8.5cm]{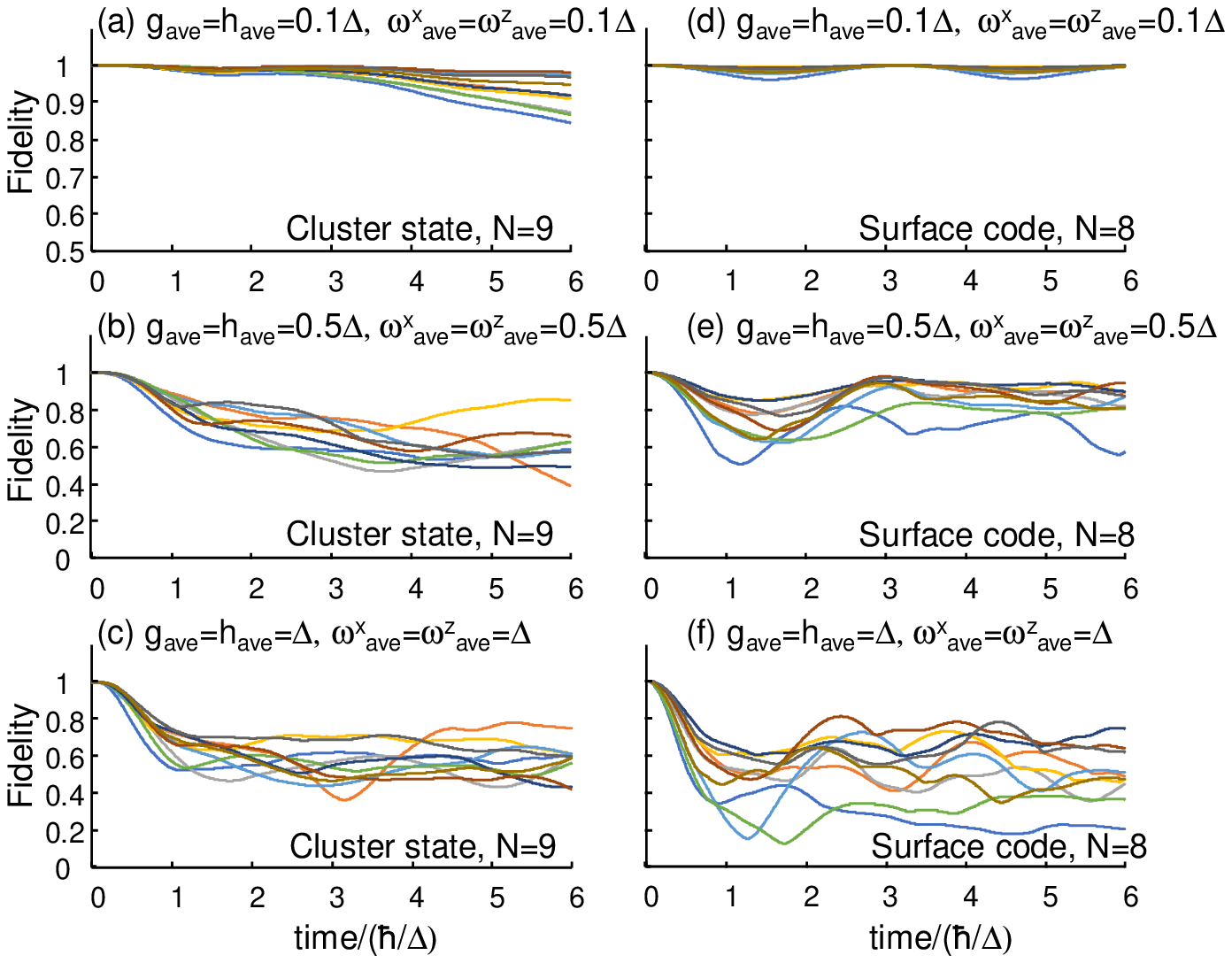}
\end{center}
\caption{Time evolution of the fidelity of the $N=9$ cluster state (a-c), and 
that of the $N=8$ surface code state (d-f) under oscillating local fields.
(a) and (d) for $g_{\rm ave}=h_{\rm ave}=0.1\Delta$,
and $\omega^x_{\rm ave}=\omega^z_{\rm ave}=0.1\Delta$. 
(b) and (e) for $g_{\rm ave}=h_{\rm ave}=0.5\Delta$,
and $\omega^x_{\rm ave}=\omega^z_{\rm ave}=0.5\Delta$. 
(c) and (f) for $g_{\rm ave}=h_{\rm ave}=\Delta$,
and $\omega^x_{\rm ave}=\omega^z_{\rm ave}=\Delta$.}
\label{FigCal4}
\end{figure}

\begin{figure}
\begin{center}
\includegraphics[width=8.5cm]{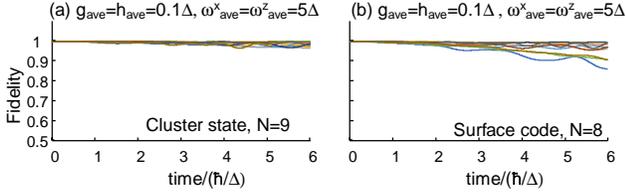}
\end{center}
\caption{Time evolution of the fidelity of the $N=9$ cluster state (a-c), and 
that of the $N=8$ surface code state (d-f) under high oscillating local fields
$\omega^x_{\rm ave}=\omega^z_{\rm ave} =5\Delta$.
(a) and (d) for $g_{\rm ave}=h_{\rm ave}=0.1\Delta$,
(b) and (e) for $g_{\rm ave}=h_{\rm ave}=0.5\Delta$, and
(c) and (f) for $g_{\rm ave}=h_{\rm ave}=\Delta$. }
\label{FigCal5}
\end{figure}

\begin{figure}
\begin{center}
\includegraphics[width=7.5cm]{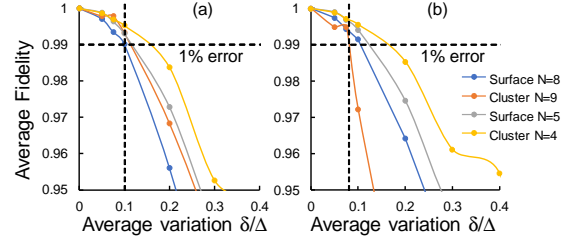}
\end{center}
\caption{Average fidelity during (a) $0<t<2\hbar/\Delta$ and
(b) $0<t<6\hbar/\Delta$ as 
a function of the average magnitude of the local fluctuations for
$\delta=g_{\rm ave}=h_{\rm ave}=\omega^x_{\rm ave}=\omega^z_{\rm ave}$.
The horizontal dashed lines show the fidelity=0.99.
The vertical dashed line shows the limit 
beyond which the quantum error correction is impossible.
}
\label{FigCal6}
\end{figure}

Figure~\ref{FigCal3} shows the numerical results of the time-dependent fidelity
for the $N=4$ cluster states (a-c) and the $N=5$ surface code states (d-f)
with oscillating fluctuations 
for finite $\omega^x_{\rm ave}=\omega^z_{\rm ave}$. 
Figure \ref{FigCal4} shows the numerical results of the time-dependent fidelity
for the $N=9$ cluster states (a-c) and the $N=8$ surface code states (d-f)
with oscillating fluctuations by finite $\omega^x_{\rm ave}=\omega^z_{\rm ave}$. 
When these figures are compared with those of 
the $N=4$ cluster states and the $N=5$ surface code states
with static local fields (Figs.~\ref{FigCal1} and \ref{FigCal2}),
there is no notable difference. 
Thus, we are led to conclude that the effect of the local variations is mainly governed 
by the magnitudes of the local fluctuations.
Figure~\ref{FigCal5} shows the time evolution of the fidelity of 
the oscillating local fields with high frequencies 
$\omega^x_{\rm ave}=\omega^z_{\rm ave} =5\Delta$
and $g_{\rm ave}=h_{\rm ave}=0.1\Delta$.
From these results, we can again say that the degradation of the fidelity
is mainly determined by the amplitude of the local fields.

\begin{figure}
\begin{center}
\includegraphics[width=8.5cm]{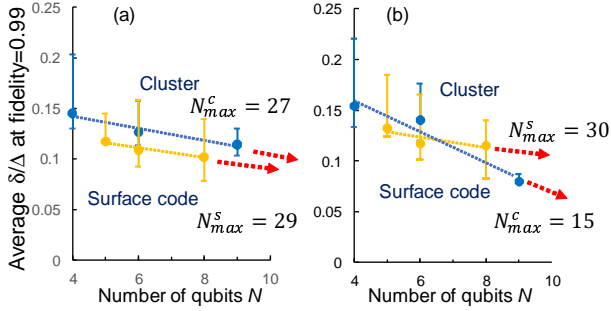}
\end{center}
\caption{
Average variation $\delta/\Delta$ at which 
the fidelity is 0.99 in Fig.~\ref{FigCal6} for the $N=9$, $N=6$, and $N=4$ surface-code states, 
and the $N=8$, $N=6$ and $N=5$ cluster states. 
The data in (a) and (b) are taken from Fig.~\ref{FigCal6}(a) and Fig.~\ref{FigCal6}(b), respectively. 
The error bars are calculated from the standard deviation of $\delta/\Delta$.
We have obtained similar results of the time evolutions of the fidelity for $N=6$ cases. 
The extrapolation equations for the cluster states and the surface-code states in (a) 
are given by $\delta/\Delta=-0.0061N + 0.1676$ 
and $\delta/\Delta=-=-0.0049N + 0.1405$, respectively.
Those in (b) are given by $\delta/\Delta= -0.0144N + 0.2156$ and 
$\delta/\Delta= -0.005N + 0.1535$, respectively.
The extrapolation of these data points to the horizontal axis 
gives a rough estimate for the maximum number of qubits $N_{max}^s$ and $N_{max}^c$ whose errors can be corrected.
}
\label{FigCal7}
\end{figure}

Next, we estimate the magnitudes 
of small local fields below which quantum error-correction can be effective.
Wang {\it et al.}~\cite{Fowler2} show that a 1 \% error can be corrected by the quantum error correction.
As mentioned above, our numerical results show that 
if the magnitude of local fluctuations are less than $0.1\Delta$, 
the fidelity remains close to unity.
To examine the possibility of the quantum error correction, 
we replot the average fidelity over a long time period compared with $\hbar/\Delta$, 
as a function of the average magnitude of the local fields, $\delta$:
\begin{equation}
\delta\equiv g_{\rm ave}=h_{\rm ave}=\omega^x_{\rm ave}=\omega^z_{\rm ave}.
\end{equation}
Figure~\ref{FigCal6} shows the average fidelity during (a) $0<t<2\hbar/\Delta$
and (b) $0<t<6\hbar/\Delta$ for 
$N=9$ and $N=4$ cluster states and for $N=8$ and $N=5$ surface code states.
For all cases, we can see some critical magnitudes around $\delta \sim 0.1\Delta$
beyond which the average fidelity rapidly decreases.
In other word, both the cluster states and the surface code states 
are robust against small local fluctuations to the extent that 
quantum error correction can be carried out.
In Fig.~\ref{FigCal6}, we can also see that 
as the number of qubits $N$ increases, the fidelity decreases
faster for cluster states than surface-code states. 
To see this scaling effect more clearly, 
we choose the variation $\delta/\Delta$ at which 
the fidelity is 0.99 in Fig.~\ref{FigCal6} and plot them as 
a function of the number of qubits in Fig.~\ref{FigCal7}.
Although the number of the qubits in the cluster states 
is not the same as that of the surface code states, 
we think that we can see a general trend by this scaling. 
The results with approximated linear equations are added in Fig.~\ref{FigCal7}.  
The cross points of these equations to the horizontal axis 
show the maximum number of qubits whose errors can be corrected.
From Fig.~\ref{FigCal7}(a), we have $N_{\rm max}^c=27.4$ 
for the cluster state and $N_{\rm max}^s=28.7$ for the surface code state.
From Fig.~\ref{FigCal7}(a), we have $N_{\rm max}^c=14.5$ 
for the cluster state and $N_{\rm max}^s=30.4$ for the surface code state.
These simple estimates will be a guide to our consideration about how much 
local fluctuations should be suppressed in order to achieve an intended qubit size.
From Ref.~\cite{Fowler}, $d \ge 5$ is required to correct errors, 
where $d$ is the surface code $distance$.
The surface code of the distance $d$ includes $d^2$ data qubits.
Our results indicate that the $d=5$ surface code is the only 
code in which the error correction is meaningful when all qubits are 
affected by their local fluctuations over a time interval of the order of 
$2\hbar/\Delta$.
However, as can be seen from Figs.~\ref{FigCal1}-\ref{FigCal5}, 
the fidelity rapidly improves for $t\ll 2\hbar/\Delta$ .
Thus, in the surface code for $d\ge 6$, the error-correction should be carried out 
during shorter time than $\hbar/\Delta$. 
The situation can be the same for the cluster state of $N \ge 24$.
Of course, we need more data to elaborate how many qubits are tolerable 
for the quantum error-correction in the future.

\begin{figure}
\begin{center}
\includegraphics[width=8.5cm]{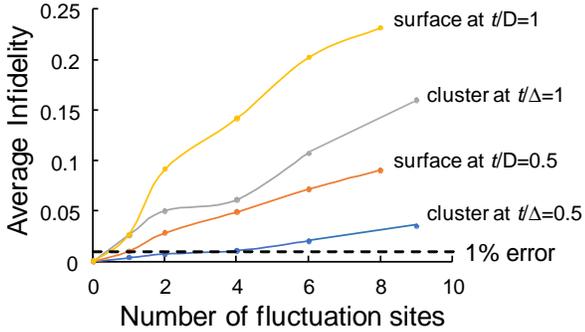}
\end{center}
\caption{
Infidelity of the $N = 9$ cluster state, and that
of the $N = 8$ surface code state under oscillating local fields 
$\omega_{\rm ave}^x =\omega_{\rm ave}^z= 0.5\Delta$ and 
$g_{\rm ave}= h_{\rm ave} = 0.5\Delta$, as a function of the
number of fluctuation sites. These results are obtained
after averaging over ten calculations
}
\label{FigCal8}
\end{figure}

Thus far, the number of the fluctuating sites are the same as 
that of qubits, that is, all qubits are assumed to be affected by local fluctuating fields.
This is the case where the active qubit area is larger than the defect-free area 
shown in Table 1.
Next, we investigate the relationship between the fidelity and 
the number of the fluctuating sites 
by assuming that we can make qubits smaller and smaller.
Figure~\ref{FigCal8} shows the infidelity 
as a function of the defect sites $N_{\rm defect}$. 
The numerical results shown above correspond to those of $N_{\rm defect}=N$.
From Fig.~\ref{FigCal8}, the infidelity is approximately proportional to 
$N_{\rm defect}$. 
That is, the reliability of the system linearly depends on the density of 
local fluctuations.
One might expect that a single local fluctuating field drastically 
changes the fidelity, because both the cluster state and 
the surface code state are highly entangled. 
However, our numerical finding that the degradation of the fidelity
is proportional to the number of fluctuating fields 
shows that the fidelity can be improved 
if the number of fluctuating sites is reduced.

Combining with the results of Figs.~\ref{FigCal7}, 
in order to construct a large qubit system whose 
number is larger than 50 to gain quantum advantage,
we have to fine-tune local fluctuations one by one 
or reduce the number of defects.
In the former case, an extra overhead of control circuits is inevitable.
In the latter case, we should reduce the area of devices.
From Table 1, if we construct a qubit system based on Si/SiO${}_2$(Si/SiGe), 
the area of a qubit should be less than 100 nm$\times$ 100 nm
(1000 nm$\times$ 1000 nm).
Thus, like the Moore law of the conventional digital circuits, 
smaller devices are better.

\section{Effect of dissipation}\label{sec:optical}
\begin{figure} 
\begin{center}
\includegraphics[width=7.0cm]{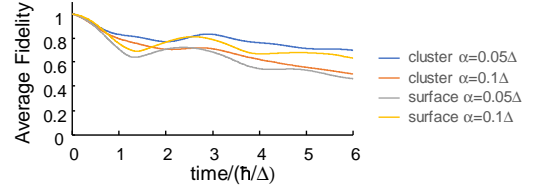}
\end{center}
\caption{Time evolution of the fidelity of the $N=4$ cluster state  
and that of the $N=5$ surface-code state under oscillating local fields 
and optical potentials $\alpha$.
The results are the average over 10 samples with random 
local fields where
$g_{\rm ave}=h_{\rm ave}=0.5\Delta$ 
and $\omega^x_{\rm ave}=\omega^z_{\rm ave}=0.5\Delta$. }
\label{FigCal9}
\end{figure}

We can use an optical potential to describe the effect of dissipation phenomenologically~\cite{Zohta}.
The Hamiltonian of the optical potential is given $H_{d}\equiv i\alpha$,
where $\alpha$ is related to the lifetime $\tau$ of the system given by  
$\alpha=1/\tau$.
By the optical potential, the fidelity decrease exponentially. 
Figure~\ref{FigCal9} shows time evolution of the fidelity 
in the presence of an optical potential. 
The fidelity decreases similarly for the cluster state and the surface code state.

\section{Discussion}\label{sec:discussion}
In this paper, we assume that the cluster-state Hamiltonian $H_0^c$ and the surface code Hamiltonian $H_0^s$ are given. 
However, the construction of the target Hamiltonian is also an important problem.
For solid-state systems, 
the most natural interaction between qubits is two-body. 
On the other hand, both the cluster-state Hamiltonian and 
the surface-code Hamiltonian includes more than three-body interactions.
Thus, we have to artificially construct the corresponding Hamiltonians
to keep those states as eigenstates.
In Ref.~\cite{tanaPRA}, we have 
shown how to derive the cluster-state Hamiltonian and 
the surface-code Hamiltonian starting from 
the general Hamiltonian of two-body interactions.
In those methods, there remain residual terms 
other than target interactions.
Thus, when we construct the target Hamiltonian, 
we will have to include many interaction terms other than $H_1$ in this paper.
The investigation of the effects of these terms will be a future problem.

Let us estimate the time scale in this paper. 
When we construct the surface-code Hamiltonian on the basis of the method of Ref.~\cite{tanaPRA},  
$\Delta$ can be estimated by using the
original coupling strength between two qubits.
From Ref.~\cite{Yoshihara}, if we use 10 MHz $< \Delta <$ 100GHz, 
the time scale expressed by $\sim \Delta^{-1}$ 
is in the range of 10ps-100us.
Thus, depending on the measurement time, the effect of 
the fluctuations described here should be able to be detected.

\section{Conclusion}\label{sec:conclusion}
We have numerically investigated the effect of random local fields on cluster states and surface code states.
We have estimated how the fidelity is affected depending on the magnitude of local fluctuations for both cluster states and surface code states.
We have shown that time evolution of the fidelity looks similar 
between the cluster states and the surface code states.
For small number of qubits ($N<10$), we 
find that the effect of the local fluctuations rapidly decreases 
to reach the 1 \% error 
when the fluctuation magnitude is 10\% of the energy gap $\Delta$
for both the cluster states and the surface code states when all qubits are subject to fluctuating fields.
We also find that, if the magnitude of fluctuations exceeds $\Delta/2$,
the fidelity for both entangled states deteriorates dramatically.
Although the number of data is small, 
it is found that the maximum number of qubits  
that can be corrected (1\% error threshold) is estimated to be less than 31 
for surface code states and 27 for cluster states, 
when the fidelity is averaged during $t<2\hbar/\Delta$.
This means that when there are local fluctuations the 
error-correction should be carried out during a time shorter than $\hbar/\Delta$.
We can also improve the fidelity by reducing the number of qubits that are subject to local field fluctuations.
More elaborate calculations will be needed to estimate the maximum number 
of qubits during a long time of the order of $\hbar/\Delta$.
The quantum advantage requires at least 50 qubits~\cite{Boixo,Villalonga} and 
our results suggest that local fluctuations 
place severe restrictions to go way above this threshold.


\end{document}